\documentclass[ %preprint,
 superscriptaddress,
 amsmath,amssymb,
 aps,
 twocolumn,
 pra,
]{revtex4-2}

\usepackage{xcolor}
\usepackage{comment}
\usepackage{graphicx}% Include figure files
\usepackage{dcolumn}% Align table columns on decimal point
\usepackage{bm}% bold math
\usepackage{hyperref}% add hypertext capabilities
\usepackage{cancel}
\begin{document}

\preprint{APS/123-QED}

\title{Fully Programmable Spatial Photonic Ising Machine by Focal Plane Division}

\author{Daniele Veraldi}
\email{daniele.veraldi@uniroma1.it}
\affiliation{Department of Physics, Sapienza University, 00185 Rome, Italy}
\author{Davide Pierangeli}
\email{davide.pierangeli@roma1.infn.it}
\affiliation{Institute for Complex Systems, National Research Council, 00185 Rome, Italy}
\author{Silvia Gentilini}
\affiliation{Institute for Complex Systems, National Research Council, 00185 Rome, Italy}
\author{Marcello Calvanese Strinati}
\affiliation{Enrico Fermi Research Center (CREF), Via Panisperna 89a, 00184 Rome, Italy}
\author{Jason Sakellariou}
\affiliation{QUBITECH, Thessalias 8, GR 15231 Chalandri, Athens, Greece}
\author{James S. Cummins}
\affiliation{Department of Applied Mathematics and Theoretical Physics, University of Cambridge, Wilberforce Road, Cambridge, CB3 0WA, United Kingdom}
\author{Airat Kamaletdinov}
\affiliation{Department of Applied Mathematics and Theoretical Physics, University of Cambridge, Wilberforce Road, Cambridge, CB3 0WA, United Kingdom}
\author{Marvin Syed}
\affiliation{Department of Applied Mathematics and Theoretical Physics, University of Cambridge, Wilberforce Road, Cambridge, CB3 0WA, United Kingdom}
\author{Richard Zhipeng Wang}
\affiliation{Department of Applied Mathematics and Theoretical Physics, University of Cambridge, Wilberforce Road, Cambridge, CB3 0WA, United Kingdom}
\author{Natalia G. Berloff} 
\affiliation{Department of Applied Mathematics and Theoretical Physics, University of Cambridge, Wilberforce Road, Cambridge, CB3 0WA, United Kingdom}
\author{Dimitrios Karanikolopoulos}
\affiliation{Key Laboratory for Quantum Materials of Zhejiang Province, Physics Department, Westlake University, 18 Shilongshan Rd, Hangzhou 310024, Zhejiang, China}
\affiliation{Institute of Natural Sciences, WIAS, 18 Shilongshan Road, Hangzhou, Zhejiang Province 310024, China}
\author{Pavlos G. Savvidis}
\affiliation{Institute of Electronic Structure and Laser, FORTH, 70013 Heraklion, Crete, Greece}
\affiliation{Key Laboratory for Quantum Materials of Zhejiang Province, Physics Department, Westlake University, 18 Shilongshan Rd, Hangzhou 310024, Zhejiang, China}
\affiliation{Institute of Natural Sciences, WIAS, 18 Shilongshan Road, Hangzhou, Zhejiang Province 310024, China}
\author{Claudio Conti}
\email{claudio.conti@uniroma1.it}
\affiliation{Department of Physics, Sapienza University, 00185 Rome, Italy}
\affiliation{Enrico Fermi Research Center (CREF), Via Panisperna 89a, 00184 Rome, Italy}
%\affiliation{Institute for Complex Systems, National Research Council, 00185 Rome, Italy}
\date{\today}

\begin{abstract}
Ising machines are an emerging class of hardware that promises ultrafast and energy-efficient solutions to NP-hard combinatorial optimization problems. Spatial photonic Ising machines (SPIMs) exploit optical computing in free space
to accelerate the computation, showcasing parallelism, scalability, and low power consumption. 
However, current SPIMs can implement only a restricted class of problems. 
This partial programmability is a critical limitation that hampers their benchmark.
Achieving full programmability of the device while preserving its scalability is an open challenge.
Here, we report a fully programmable SPIM achieved through a novel operation method based on the division of the focal plane. In our scheme, a general Ising problem is decomposed into a set of Mattis Hamiltonians, whose energies are simultaneously computed optically by measuring the intensity on different regions of the camera sensor. Exploiting this concept, we experimentally demonstrate the computation with high success probability of ground-state solutions of up to 32-spin Ising models on unweighted maximum cut graphs with and without ferromagnetic bias.
Simulations of the hardware prove a favorable scaling of the accuracy with the number of spin. 
Our fully programmable SPIM enables the implementation of many quadratic unconstrained binary optimization problems, further establishing SPIMs as a leading paradigm in non von Neumann hardware.
\end{abstract}
\maketitle

Ising machines (IMs) are specialized devices designed to solve quadratic unconstrained binary optimization (QUBO) problems by finding the ground state of the corresponding Ising model~\cite{McMahon2022}.
IMs harness physical effects exhibited by the underlying system as a mechanism to accelerate the ground state search.
Spatial photonic Ising machines (SPIMs) encode Ising spins in the optical phase and
exploit spatial light modulation and coherent optical propagation in free space to compute optically the value of the Ising Hamiltonian~\cite{Pierangeli2019}.
Taking advantage of the spatial parallelism~\cite{McMahon2023}, as well as the high resolution of spatial light modulators (SLMs) and low-intensity continuous-wave lasers, SPIMs showcased parallel operation, energy efficiency, and scalability~\cite{Pierangeli2019}. A key issue concerns their programmability, i.e.,
the capability to program the spin interaction matrix to realize any QUBO problem. 
The first SPIMs could not map graphs with arbitrary connections.
This partial programmability restricts the class of problems that can be implemented~\cite{Wang2024}.
Recently, many approaches have been pushed forward to extend the range of applicability of SPIMs~\cite{Pierangeli2021,
Pierangeli2020_1, Pierangeli2020_2, Leonetti2021, Jacucci2022, Kumar2020, Kumar2023, Ruan2021_1, Ruan2021_2, Zhang2022, Feng2024, Zhang2024, Suzuki2023, Ruan2023}.
Among them, Z. Ruan and co-authors~\cite{Ruan2021_1, Ruan2021_2} developed the so-called Gauge method that allows
a simple programming of the so-called Mattis Hamiltonian (rank-$1$ interaction matrices).
By exploiting the Gauge encoding, a SPIM implementing full-rank coupling matrices has been demonstrated through 
their decomposition over multiple wavelengths~\cite{Ruan2023}.
However, this wavelength-division multiplexing scheme requires as many wavelengths as spins and suffers from chromatic dispersion, factors that limit its scalability. Achieving full programmability in SPIMs while maintaining scalability remains an open challenge.

In this Letter, we present and experimentally demonstrate a new SPIM based on focal plane division (FPD)
to solve any Ising model while preserving scalability.
The scheme uses a single SLM in a novel configuration to
decompose any Ising problem into a set of Mattis Hamiltonians that are computed in parallel
by dividing and measuring the optical intensity on separate regions of the camera sensor.
This new arrangement of the setup enhances the computing capability of SPIMs by enabling the exploitation
of the additional spatial degrees of freedom of the focal plane.
We experimentally demonstrate ground state solutions of different types of graphs:
M\"obius ladder, Max-Cut, and Max-Cut with a ferromagnetic (FM) bias with 16 and 32 spins, 
with a success probability of $95\%$, $50\%$, $55\%$, and $10\%$, respectively. 
We compare simulations of the hardware with simulated annealing (SA), 
showing that the device accuracy scales favourably the number of spins.
Our fully programmable SPIM allows photonic computation of many QUBO problems.

%%%%%%%%%%%%%%%%%%%%%%%%%%%%%%%%%%%%%%%%%%%%%%%%%%%%%%%%%%%%%%%%%%%%%%%%%%%%%%%%%%%%%%%%%%%%%%%%%%%%%%%%%%%%%%%%%%%%%
\begin{figure}
\centering
\hspace*{-1.6cm}
\includegraphics[width = 11.7 cm, keepaspectratio]{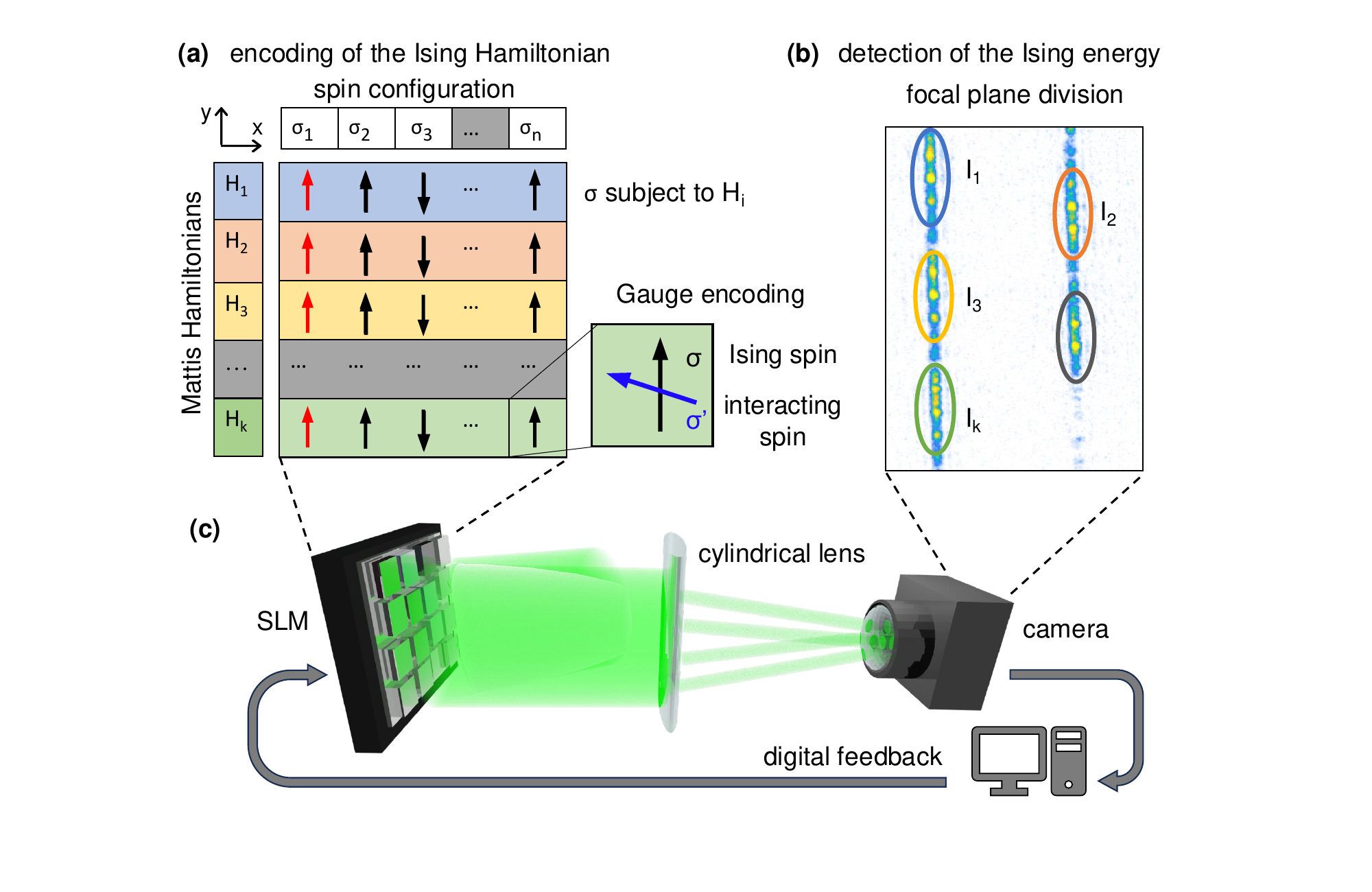}
\vspace*{-1cm}
\caption{SPIM by focal plane division. (a) Schematic of the SLM screen division. The same spin configuration 
$\bm{\sigma}$ is encoded in each row. We apply the Gauge encoding to implement each Mattis Hamiltonian 
$\mathcal{H}_k$ on each row. (b) Intensity detected in the focal plane. By using blazed gratings,
we perform a division of the focal plane to separate the signals of different $\mathcal{H}_k$.
(c) Sketch of the experimental setup. The spin energy is evaluated by the measured intensities $[I_1,...,I_K]$. The ground-state search operates by updating via digital feedback $\bm{\sigma}$ on the SLM 
according to the measured energy and a SA algorithm.
  \label{fig:concept} }
\end{figure}
%%%%%%%%%%%%%%%%%%%%%%%%%%%%%%%%%%%%%%%%%%%%%%%%%%%%%%%%%%%%%%%%%%%%%%%%%%%%%%%%%%%%%%%%%%%%%%%%%%%%%%%%%%%%%%%%%%%%%

The Ising model is defined by the Hamiltonian
\begin{eqnarray}
    H\left(\bm{\sigma}\right) = -\frac{1}{2}\bm{\sigma}^T \bm{J\sigma}
\end{eqnarray}
with spin configuration $\bm{\sigma}= \left[ \sigma_1,...,\sigma_N\right]$ and $\sigma_i=\pm1$, 
$N$ the number of spins, and symmetric real-valued interaction matrix $\bm{J}$.
A fully programmable IM thus requires $N^2$ programmable units.
Most SPIMs~\cite{Pierangeli2020_2, Ruan2021_1} realized and solved Eq.~(1) with rank$(\bm{J})=1$.
These Mattis-type interactions are characterized by the outer product $ \bm{J} = \bm{\xi\xi}^T$.
We decompose the interaction matrix of a general Ising problem into a linear combination of Mattis problems 
$ \bm{J} = \sum_{k=1}^K\lambda_k\bm{\xi}_k\bm{\xi}_k^T $, so that
\begin{eqnarray}
    H\left(\bm{\sigma}\right) = -\sum_{k=1}^K\lambda_k \mathcal{H}_k\left(\bm{\sigma}\right) = -\frac{1}{2}\sum_{k=1}^K \sum_{i,j=1}^N\lambda_k \xi_{ik}\xi_{jk}\sigma_i\sigma_j
\label{hamiltonian_mattis}
\end{eqnarray}
Therefore, a SPIM capable to implement the Mattis Hamiltonian $\mathcal{H}_k$
can process a full-rank Ising problem ($K = N$)
when operating through a multiplexing scheme~\cite{Zhang2024}.
To encode each $\mathcal{H}_k$ in the optical field, we exploit phase-only encoding by the Gauge transformation method~\cite{Ruan2021_1, Ruan2021_2}. Through this transformation we pass from binary to circular spins that encode also the interaction by continuous phase values.
The coupling coefficients $-1\leq\xi_{ik}\leq+1$ are encoded by introducing a rotation of the Ising spin.
The Gauge map rotates each spin $\sigma_i$ by an angle $\alpha_{ik}=\arccos(\xi_{ik})$ 
with respect to the z-axis to obtain the spin vector $\bm{\sigma}^\prime$. 
The z-component $\sigma_{ik}^{\prime z}=\xi_{ik}\sigma_i$ is the effective spin that is encoded.
The Mattis Hamiltonian remains invariant under this transformation~\cite{Ruan2021_1},
allowing to implement $\mathcal{H}_k$ by using the sole optical phase.

To experimentally realize a fully programmable SPIM, we design a FPD method to compute simultaneously all the
$\mathcal{H}_k$ by a single-shot intensity measurement. We apply an eigen-decomposition to the target $\bm{J}$ and use the eigenvectors to define the coupling strengths $\lambda_k$ in Eq.~(2).
As illustrated in Fig.~\ref{fig:concept}(a), to optically compute in parallel all the $\mathcal{H}_k$,
we divide the SLM screen in $K$ rows. Each row contains the same Ising spin configuration $\bm{\sigma}$.
We then apply a different $\mathcal{H}_k$ to each row by Gauge encoding, 
i.e. we encode the $i$-th effective spin of the $k$-th row according to $\xi_{ik}$.
Ising spins are mapped into phase delays $\pm \pi/2$.
On the SLM, each spin corresponds to a macropixel made by $c=P_x \times P_y$ pixels
that is phase modulated as
\begin{eqnarray}\label{eq.phase}
    \phi_{ik}^{l} = \sigma_i\frac{\pi}{2} + (-1)^l\alpha_{ik},
\end{eqnarray}
being $\sigma_{ik}^{\prime z}=\exp{(i \phi_{ik}^{l})}$, 
where $l$ is the pixel index along the $x$-axis of the SLM ($1<l<P_x$) 
and the rotation angle $\alpha_{ik}$ is applied to the $i$-th spin of the $k$-th row.
Equation~\eqref{eq.phase} is derived in Refs. ~\cite{Ruan2021_1}.
Each macropixel thus encodes information both on the spin and its interactions.
The division by rows is an optimal method to exploit the SLM plane. %for optical computing.
Specifically, we are programming $N^2$ coefficients by a single SLM, in analogy with optimal schemes 
for optical vector-matrix multiplications~\cite{Spall2020}.

To measure the optical intensity associated to each $\mathcal{H}_k$, 
we analyze by a camera the intensity distribution of the propagated beam in the Fourier plane.
Fig.\ref{fig:concept}(b) reports an experimental image showing the intensity on the focal plane.
We use a combination of a cylindrical lens and digital blazed gratings to ensure no mixing between the signals of different rows and to spatially separate the intensities $I_k$ along the $y$-axis of the camera sensor.
Each $I_k$ gives a measurement of the energy of $\mathcal{H}_k$ for a given $\bm{\sigma}$.
Note that $I_k$ presents multiple internal spots [Fig.\ref{fig:concept}(b)],
which result from diffraction from the periodic components of the SLM phasemask. 
Therefore, we first identify the position of the $k$-th focal spot (intensity maximum)
and then integrate the signal within a rectangular region of interest ($3\times 50$ camera pixels) 
to accurately measure the set of $I_k$. 
The optical Ising energy is evaluated as $F=\sum_k \lambda_k I_k$.

%%%%%%%%%%%%%%%%%%%%%%%%%%%%%%%%%%%%%%%%%%%%%%%%%%%%%%%%%%%%%%%%%%%%%%%%%%%%%%%%%%%%%%%%%%%%%%%%%%%%%%%%%%%%%%
\begin{figure*}
\vspace*{0.1 cm}
\hspace*{-0.2cm}
\includegraphics[width = 18.2 cm, keepaspectratio]{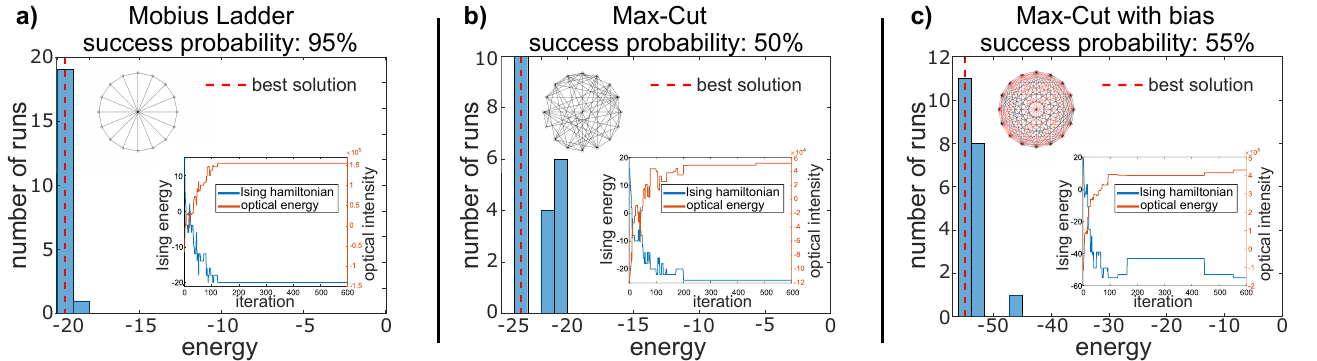}
%\vspace*{-0.2cm}
\caption{Ground-state accuracy of the SPIM by FPD with $N=16$ for (a) M\"obius ladder, (b) Max-Cut and (c) Max-Cut with ferromagnetic bias. The energy histograms show a ground state probability of $95\%$, $50\%$ and $55\%$, respectively.
Insets shows the optical energy $F$ and the value of the Ising Hamiltonian during an experimental run. \label{fig:experiment_results}}
\end{figure*}
%%%%%%%%%%%%%%%%%%%%%%%%%%%%%%%%%%%%%%%%%%%%%%%%%%%%%%%%%%%%%%%%%%%%%%%%%%%%%%%%%%%%%%%%%%%%%%%%%%%%%%%%%%%%%%

%%%%%%%%%%%%%%%%%%%%%%%%%%%%%%%%%%%%%%%%%%%%%%%%%%%%%%%%%%%%%%%%%%%%%%%%%%%%%%%%%%%%%%%%%%%%%%%%%%%%%%%%%%%%%%
\begin{figure}[b]
\includegraphics[width = 6 cm, keepaspectratio]{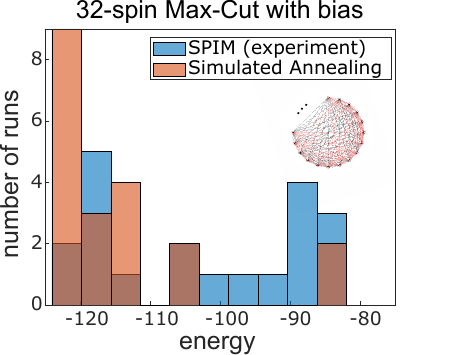}
\vspace*{-0.1cm}
\caption{Accuracy of the SPIM by FPD for $N=32$ on fully connected biased Max-Cut graphs.
Energy distribution of the experimental and SA solutions ($P_{\rm{suc}}=10\%$).
}
\label{fig:experiment32spin}
\end{figure}
%%%%%%%%%%%%%%%%%%%%%%%%%%%%%%%%%%%%%%%%%%%%%%%%%%%%%%%%%%%%%%%%%%%%%%%%%%%%%%%%%%%%%%%%%%%%%%%%%%%%%%%%%%%%%%

The experimental device is illustrated in Fig. \ref{fig:concept}(c). 
A linearly-polarized continuous-wave $100$ mW laser at $\lambda=532$ nm is expanded on a phase-only SLM (Hamamatsu LCOS-SLM X15213S).
The phase modulation in Eq. (3) is applied by using $213$ levels of precision within the interval $[0, 2\pi]$.
The SLM operates with a measured diffraction efficiency of $0.005$.
The phase modulated beam is imaged by a cylindrical lens (focal length $f=150$ mm, numerical aperture NA$=0.1$)  on a 12-bit CMOS camera (Basler a2A2590-60umPRO). 
The total optical power impinging on the camera is $0.05$ mW.

We experimentally validate the fully programmable SPIM by evaluating its performance in finding the ground state of various Ising problems for $N = 16$.  We select a macropixel size of $18 \times 18$ pixels,
chosen to maximize the computational accuracy.
The ground state search is conducted by using digital feedback to recurrently update the spins according to a Metropolis-Hasting algorithm fed by the optical energy $F$.
During the machine run, the spin temperature is varied via a SA algorithm implemented following Ref.~\cite{Isakov2015}.
We choose three types of graphs that are widely used as benchmarks for IMs~\cite{McMahon2022}: 
M\"obius ladder, Max-Cut and Max-Cut with FM bias (biased).
M\"obius ladder is a circulant graph where every spin interacts antiferromagnetically $J_{ij} = -1$ with its nearest neighbours and the diametrically opposite spin.
For the considered Max-Cut graphs, the $J_{ij}$ are extracted randomly among $0$ and $-1$ with $0.5$ probability.
The Max-Cut with FM bias is a complete random binary graph defined by setting $J_{ij}=\pm 1$ with $0.5$ probability.
Figure \ref{fig:experiment_results} shows the computing result for each graph over $20$ machine runs with different initial conditions. Convergence is achieved on the order of $10^2$ iterations.
We compare the experimental solutions with a SA algorithm~\cite{Isakov2015}. 
The success probability $P_{\mathrm{suc}}$ is calculated with respect to the best SA solution. 
We obtain the ground state with $P_{\mathrm{suc}}$ of $95\%$, $50\%$ and $55\%$, respectively.
This performance difference reflects the fact the Möbius ladder is polynomially solvable, 
while Max-Cut graphs are classified as NP-hard~\cite{Kalinin2022, Strinati2021}. 
We remark that also the SPIM operates through SA but using the optically energy $F$. 
Both SA hyper-parameters have been fine tuned to achieve optimal results.
To ensure the agreement between the experimental energy $F$ and the corresponding Ising Hamiltonian, 
we monitor these quantities while the machine is running. As shown in Fig.\ref{fig:experiment_results},
they are anticorrelated at every iteration despite experimental noise, proving the correct operation of the SPIM during the optimization.

%%%%%%%%%%%%%%%%%%%%%%%%%%%%%%%%%%%%%%%%%%%%%%%%%%%%%%%%%%%%%%%%%%%%%%%%%%%%%%%%%%%%%%%%%%%%%%%%%%%%%%%%%%%%%%
\begin{figure*}
\includegraphics[width = 17 cm, keepaspectratio]{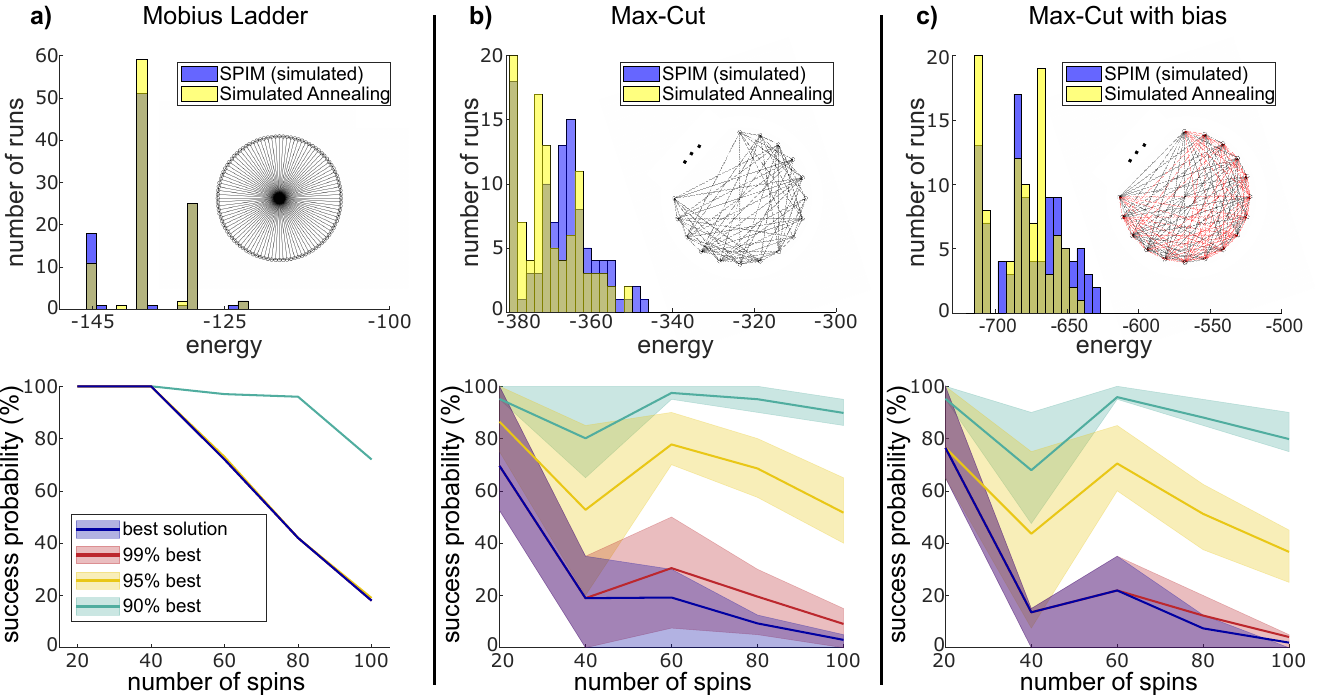}
\caption{Scaling analysis of the SPIM accuracy. Energy histograms on $100$-spin (a) M\"obius ladder, (b) Max-Cut and (c) biased Max-Cut graphs for the simulated SPIM and SA.
The probability of finding low-energy solutions within $a\%$ of the energy minimum 
is shown by increasing $N$. 
Shaded areas represent the interquartile range over $100$ different random graphs. 
\label{fig:simulation_results}}
\end{figure*}
%%%%%%%%%%%%%%%%%%%%%%%%%%%%%%%%%%%%%%%%%%%%%%%%%%%%%%%%%%%%%%%%%%%%%%%%%%%%%%%%%%%%%%%%%%%%%%%%%%%%%%%%%%%%%%

To prove the scalability of the FPD scheme, we scale up the experimental device and implement larger fully connected graphs. The macropixel size $c$ is the key factor to scale up our SPIM.
The number of SLM pixels required for general problems of size $N$ scales as $c\times N^2$,
with $c$ that is tunable within a range that depends only on the experimental setup. 
By reducing the macropixel size to $c=9\times 9$ pixels, 
we implement $32$-spin fully connected biased Max-Cut graphs ($1024$ programmable couplings).
This scale is two times larger than achieved by wavelength-multiplexed SPIMs~\cite{Ruan2023}
and competes with photonic and electronic state-of-the-art IMs that have full programmability 
without relying on digital hardware to implement the interactions~\cite{Feng2024, Kim2023}.
Figure \ref{fig:experiment32spin} reports the obtained energy histogram in comparison with SA.
We achieve $P_{\rm{suc}}=10\%$ and a good agreement between the two energy distributions. 
The performance drop is due to the lower signal-to-noise ratio on the focal plane when reducing the macropixel. This limitation can be overcome by improving the NA and dynamic range of the imaging system.

To assess further the scalability, we conduct simulations of the SPIM up to $N=100$
on the same tpye of graphs of the experiments. 
Figure \ref{fig:simulation_results} shows the energy histogram for $N = 100$ compared with SA.
For the M\"obius ladder, Max-Cut, and Max-Cut with FM bias, 
we obtain a $P_{\rm{suc}}$ of $18\%$, $18\%$ and $13\%$, respectively.
In all three cases, the accuracy of the simulated device matches the SA performance. 
This indicates that the SPIM can perform optimizations with accuracy comparable to heuristic algorithms.
Fig. \ref{fig:simulation_results} also reports $P_{\rm{suc}}$ at different $N$, 
along with the probability of finding low-energy states within $a\%$ of the energy minimum. 
For large $N$, approximate solutions are found with high probability and a favourable scaling.
We test $100$ random graphs and simulate $20$ SPIM runs for graph.
The interquartile ranges show that distinct graph instances have a different impact the convergence.
The effect, related to the hardness of Max-Cut and biased Max-cut graphs,
has been also observed in coherent IMs~\cite{McMahon2016}
on $16$-spin cubic graphs as a graph-dependent $P_{\rm{suc}}$ distribution.

We analyze the possible advantage of our SPIM over digital hardware and other IMs in terms of the expected computation time and energy efficiency. We estimate the SPIM run time as
$\tau_{\rm{run}}=\tau_{\rm{iter}} N_{\rm{iter}}=\tau_{\rm{SLM}} \alpha N$, where
$\tau_{\rm{iter}}$ is the iteration time, $N_{\rm{iter}}$ the number of iterations, 
$\tau_{\rm{SLM}}$ the SLM response time, and $\alpha$ is a parameter tunable by the annealing schedule.
The resulting time-to-solution~\cite{McMahon2022} reads as
$\tau_{\rm{sol}}= \tau_{\rm{SLM}} \alpha N \left[ \ln(0.01)/\ln(1-P_{\rm{suc}}) \right]$.
The performance of our proof-of-concept are limited by the $60$ Hz SLM frame rate ($\tau_{\rm{SLM}}\approx0.02$ s). By considering electro-optic SLMs with frame rates $>1$ GHz that are under development~\cite{Englund2024}, we get $\tau_{\rm{sol}} \sim 10 \mu$s on $100$-spin biased Max-Cut ($P_{\rm{suc}}$ from Fig. 4).
This value compares with the $\tau_{\rm{sol}} \approx 30 \mu$s of the best performing digital architectures~\cite{Goto2021}. At $N=10^{4}$, the estimated $\tau_{\rm{SLM}}$ to compete with 
specialized digital hardware becomes $\tau_{\rm{SLM}}\sim 10^{-5}$ s, i.e. only one order of magnitude 
smaller than off-the-shelf MEMS-based SLMs~\cite{Phillips2024}. %$10$ KHz
This shows the favourable scaling of $\tau_{\rm{iter}}$ of SPIMs vs digital computing, i.e. the so-called optical advantage~\cite{Pierangeli2021}.
%Since the SPIM computation time of the Ising energy is size independent, 
%there exist a critical size $N_{\rm{adv}}$ above which the SPIM features an optical advantage over any machine employing digital hardware. We estimate $N_{\rm{adv}}\sim 10^{5}$ with current SLMs technology.
When comparing with the performance of coherent IMs on biased Max-Cut~\cite{Inagaki2016, Hamerly2019},
we find that our SPIM is competitive for $\tau_{\rm{SLM}}\sim 10^{-7}$s ($10^{-5}$s) at $N=100$ ($1000$). 
%An advantage with available high-speed LCOS SLMs is expected for $N>10000$. 
As for energy efficiency, our SPIM has low power consumption: it works by using $10$mW of optical power
and overall consumes $P_{\rm{tot}}=50$ W (coherent IMs have $P_{\rm{tot}}\sim1$ kW~\cite{Takesue2021}).
We estimate the energy-to-solution as $E_{\rm{sol}}=P_{\rm{tot}}\tau_{\rm{sol}}$, 
which gives $E_{\rm{sol}}\sim 0.1$ mJ for $100$-spin Max-Cut (comparable with predictions for opto-electronic IMs \cite{VanDerSande2019, Li2021}). At this scale, the most energy-efficient IMs 
that are based on memristors~\cite{Strachan2020, MEM2023} and stochastic electronic oscillators \cite{Datta2021, Singapore2024} have a predicted $E_{\rm{sol}}\sim100$ nJ.
Importantly, while for these IMs $P_{\rm{tot}}$ will grows considerably with $N$, 
for our SPIM $P_{\rm{tot}}$ is independent of the machine scale.
This property suggests a possible advantage at a large scale also in energy efficiency.

Our fully programmable SPIM can reach large scales by shrinking further the macropixel and employing more SLM pixels. In our setup, we use a only central portion of the SLM ($300\times300$ pixels) 
to avoid aberrations and vignetting that affect the accuracy.
By exploiting the entire SLM screen via optics with larger NA, 
we can accommodate more than $100$ all-to-all connected spins already by the demonstrated macropixel size.
Available high-resolution SLMs (10M pixels) within an engineered imaging setup
would allow us to realize more than $1000$ spins readily by using modes of $c=16$ pixels, 
indicating a promising route to reach an advantage in computational performance.

In conclusion, we have demonstrated a fully programmable and scalable SPIM based on the division of focal plane. The device is simple, algorithm-agnostic, low-cost and compact by using low-power monochromatic light and phase-only modulation by a single SLM. 
Our work provides an advantageous method to optically accelerate the computation of QUBO problems.

\vspace*{0.1cm}
D.V., D.P, S.G., M.C.S., J.S., J.S.S., A.K., M.S., R.Z.W., D.K, P.G.S., N.G.B. and C.C. acknowledge support from HORIZON EIC-2022-PATHFINDERCHALLENGES-01 HEISINGBERG Project no. 101114978. D.V., D.P and C.C. acknowledge 
the National Quantum Science and Technology Institute (NQSTI) under the National Recovery and
Resilience Plan (NRRP) funded by the EU-NextGenerationEU.

\end{document}